\documentstyle[11pt]{article}
\begin{document}
\setcounter{footnote}{0}
\renewcommand{\thefootnote}{\alph{footnote}}
\renewcommand{\theequation}{\thesection.\arabic{equation}}
\newcounter{saveeqn}
\newcommand{\add}{\addtocounter{equation}{1}}
\newcommand{\alpheqn}{\setcounter{saveeqn}{\value{equation}}%
\setcounter{equation}{0}%
\renewcommand{\theequation}{\mbox{\thesection.\arabic{saveeqn}{\alph{equation}}}}}
\newcommand{\reseteqn}{\setcounter{equation}{\value{saveeqn}}%
\renewcommand{\theequation}{\thesection.\arabic{equation}}}
\newenvironment{nedalph}{\add\alpheqn\begin{eqnarray}}{\end{eqnarray}\reseteqn}
\newsavebox{\DELVECRIGHT}
\sbox{\DELVECRIGHT}{$\stackrel{\rightarrow}{\partial}$}
\newcommand{\PARVECR}{\usebox{\DELVECRIGHT}}
\thispagestyle{empty}

\begin{flushright}
IPM/P-2006/005
\par
hep-th/0602023
\end{flushright}
\vspace{0.5cm}
\begin{center}
{\large\bf{Dynamics of  $O(N)$ Model in a Strong Magnetic Background Field\\ as a Modified Noncommutative Field Theory}}\\
\vspace{1cm} {\bf Amir Jafari Salim$^{\dagger,}$}\footnote{\normalsize{Electronic address:
jafari\_amir@mehr.sharif.edu}}\hspace{0.2cm}and\hspace{0.2cm}{\bf N\'eda
Sadooghi$^{\ddagger,}$}\footnote{\normalsize{Electronic address: sadooghi@sharif.edu}} \\ \vspace{0.5cm}
{\sl ${\ ^{\dagger, \ddagger}}$Department of Physics, Sharif University of Technology}\\
{\sl P.O. Box 11365-9161, Tehran-Iran}\\
and\\
{\sl ${\ ^{\ddagger}}$Institute for Studies in Theoretical Physics and Mathematics (IPM)}\\
{\sl{School of Physics, P.O. Box 19395-5531, Tehran-Iran}}\\
\end{center}
\vspace{0cm}
\begin{center}
{\bf {Abstract}}
\end{center}
\begin{quote}
In the presence of a strong magnetic field, the effective action of a composite scalar field in an scalar $O(N)$
model is derived using two different methods. First, in the framework of worldline formalism, the 1PI $n$-point
vertex function for the composites is determined in the limit of strong magnetic field. Then, the $n$-point
effective action of the composites is calculated in the regime of lowest Landau level dominance. It is shown
that in the limit of strong magnetic field, the results coincide and an effective field theory arises which is
comparable with the conventional noncommutative field theory. In contrast to the ordinary case, however, the
UV/IR mixing is absent in this modified noncommutative field theory.
\end{quote}

\vspace{0.8cm}
\par\noindent February 2006

\newpage
\setcounter{page}{1}

\subsubsection*{1\hspace{0.3cm}Introduction}
In the recent years, there has been remarkable interest in noncommutative geometry \cite{nc}, which has made a
dramatic appearance in string theory \cite{witten} and has made noncommutative field theory (NCFT) \cite{ym} an
active field of study. The various aspects of noncommutative gauge theories have been extensively studied and a
number of novel phenomena discovered (for a review see  \cite{DNS} and the references therein). The conventional
noncommutative gauge theory is characterized by replacing the familiar product of functions by the Moyal
$\star$-product and is therefore a nonlocal theory involving higher order derivatives between the fields. From
perturbative point of view, the theory consists therefore of planar and nonplanar diagrams. The latter are
usually the source of the appearance of a certain duality between ultraviolet (UV) and infrared (IR) behavior of
the theory. This UV/IR mixing phenomenon manifests itself in the singularity of the amplitudes in two limits of
small noncommutativity parameter $\theta$ and large cutoff $\Lambda$ of the theory \cite{MRS}.
\par
Apart from these features noncommutative field theory exhibits certain dynamics of a quantum mechanical model in
a strong magnetic field \cite{Bigatti}. Recently the connection between the dynamics in relativistic field
theories in a strong magnetic background field and that in NCFT has been studied in \cite{EV,  ESV, EMV, Hash}.
In \cite{EV}, the nontrivial dynamics of the fermionic Nambu-Jona-Lasinio (NJL) model in a constant magnetic
background is considered. The effective action of this theory is determined for strong magnetic field in the
regime of lowest Landau level (LLL) dominance and its dynamics is compared with a conventional noncommutative
field theory. Similarly the chiral dynamics of QED and QCD is shown to be governed by a complicated nonlocal
NCFT \cite{ESV, EMV, Hash}. In all this cases, however, the emergent effective noncommutative field theory is
different from the NCFT ones considered in the literature \cite{ym}. In particular, the UV/IR mixing \cite{MRS},
taking place in the conventional NCFT is absent in these class of ``modified'' noncommutative field theories.
\par
In this paper, we will present another example of this phenomenon. Here, we will determine the effective action
of a composite scalar field in a scalar $O(N)$ model in the presence of a strong magnetic background field using
two different methods. In section 2, we will introduce the scalar $O(N)$ model in Euclidean spacetime and will
compute the effective action for the composite scalar field. In section 3, in the framework of worldline path
integral method \cite{BK, Strassler, focksch, worev} the 1PI $n$-point vertex function for the composite field
will be determined in the limit of strong magnetic background field. The emergent $n$-point function will be
comparable with the $n$-point vertex function of a certain modified noncommutative field theory, where no UV/IR
mixing occurs. In this case, as in the previous cases \cite{EV, ESV, EMV, Hash}, besides the usual Moyal phase
factor, an additional Gaussian like form factor appears in the $n$-point vertex function of the composites. This
exponentially damping form factor reflects an inner structure of the composites and is responsible for the
removal of the UV/IR mixing.
\par
In section 4, the effective action of $n$ composite fields will be calculated in the strong magnetic field
limit, using the same method as in \cite{EV}. First the Green's function of the theory will be determined in the
regime of LLL dominance. Then, using this effective propagator the contribution of $n$ composites to the full
effective action of the theory will be determined. In section 5, the 1PI $n$-point vertex from section 3 will be
compared with the LLL $n$-point effective action from section 4. We will show that here, as in NJL model in the
presence of constant magnetic background \cite{Gusynin}, a dimensional reduction from $D=4$ to $D=2$ occurs in
the longitudinal section of the effective theory where a free propagation of the composites are observed. In the
transverse section however, a modified noncommutative field theory arises where no UV/IR mixing occurs.
\subsubsection*{2\hspace{0.3cm}Effective Action for Composite Scalar Field}
\setcounter{section}{2} \setcounter{equation}{0}
Let us start with the Lagrangian density of a scalar $O(N)$
model in Euclidean spacetime
\begin{eqnarray}\label{S1-1}
{\cal L}=-| D_\mu\Phi|^{2}-m^{2}{\Phi}^{\ast}\Phi -\lambda({\Phi}^{\ast}\Phi)^2,
\end{eqnarray}
where $\Phi=(\phi_1,\phi_2,\cdots,\phi_N)$  and covariant derivative $D_\mu$ is defined by
\begin{eqnarray}
D_\mu\Phi={\partial}_\mu\Phi+ie A_\mu\Phi.
\end{eqnarray}
To study the dynamics of the bound state formed in this theory\footnote{In appendix A, we will show that even in
a theory with one fermion flavor a condensate is built in the presence of a strong magnetic field.}, we
introduce the composite field $\sigma \equiv\lambda{\Phi}^{\ast}\Phi$ and a new coupling constant
$g\equiv\lambda N$. The Lagrangian density (\ref{S1-1}) can then be given by
\begin{eqnarray}
{\cal L}=-| D_\mu\Phi|^{2}-m^{2}{\Phi}^{\ast}\Phi -\sigma{\Phi}^{\ast}\Phi+\frac{N}{2g}\sigma^2,
\end{eqnarray}
and the effective action for $\sigma$ reads
\begin{eqnarray}
\tilde{ \Gamma}[\sigma]=\Gamma[\sigma]+\frac{N}{2g}\int d^4 x\ \sigma^2,
\end{eqnarray}
where $\Gamma[\sigma]$ is found using the standard deformation
\begin{eqnarray}
e^{\Gamma[\sigma]}&=&\int\ {\cal{D}}\Phi^\ast\ {\cal{D}}\Phi\ \exp\left( \int d^4x\  [|
D_\mu\Phi|^{2}+m^{2}{\Phi}^{\ast}\Phi +\sigma{\Phi}^{\ast}\Phi] \right)\nonumber\\
&=& \int\ {\cal{D}}\Phi^\ast\ {\cal{D}}\Phi\ \exp\left( \int d^4 x\ {\Phi}^{\ast}[- D_\mu D^\mu+m^{2} +\sigma]
\Phi \right)\nonumber\\
&=& \exp\left(-{\rm Tr }\ln[-D_\mu D^\mu+m^{2} +\sigma]\right),
\end{eqnarray}
and is therefore given by
\begin{eqnarray}\label{xx}
\Gamma[\sigma]= -{\rm Tr }\ln[-D_\mu D^\mu+m^{2} +\sigma].
\end{eqnarray}
In the following two sections, we will calculate $\Gamma[\sigma]$ in the presence of a strong background
magnetic field using two different methods. In section 3, we will calculate the 1PI $n$-point vertex function
$\Gamma_{\mbox{\small{1PI}}}[p_{1},\cdots,p_{n}]$ using the worldline path integral formalism \cite{BK,
Strassler, focksch, worev}. In section 4, we will follow the method which was used in \cite{EV} to determine the
effective action of a $(D+1)$-dimensional Nambu-Jona-Lasinio (NJL) model in the regime of lowest Landau level
(LLL) dominance.
\subsubsection*{3\hspace{0.3cm}Effective Action of the Composite Field from the Worldline Formalism}
\setcounter{section}{3} \setcounter{equation}{0}
\par
The worldline formalism is originally introduced in \cite{BK} as a useful mathematical tool to study the
effective field theory limit of the underlying string theory. In this formalism, all path integrals are
manipulated into Gaussian form, and this reduces their computation to the calculation of worldline propagators
and determinants. In this section, we will keep very closely to the notation of \cite{worev},\footnote{The
reader is also referred to the excellent review of M. Strassler \cite{Strassler} for more useful examples and
details.} where among many other examples the $n$-point amplitude of a massive $\varphi^{3}$ theory in the
one-loop level and QED in a constant magnetic background field are calculated separately. Here, we will combine
the results presented in \cite{worev}, and will determine the $n$-point vertex function of the composite field
$\sigma$. At the end, we will consider the limit of a constant but strong magnetic field as background. In this
way, we arrive at an effective vertex function which is comparable with the vertex function of a certain
modified noncommutative field theory (NCFT) for the composite field $\sigma$, where besides the Moyal
$\star$-product of the conventional noncommutative field theory, a certain exponentially damping factor occurs
in the interaction vertices of $\sigma$ \cite{EV}. This factor is shown to play an important role in providing
consistency of this class of modified NCFT \cite{EV, ESV, EMV, Hash}.
\par
Let us start with the worldline path integral representation for the effective action (\ref{xx}), which reads
\cite{worev, Strassler}
\begin{eqnarray}\label{rawGamma}
\hspace{-1cm}\Gamma[\sigma]=N \int_{0}^{\infty}\frac{dT}{T} e^{-m^{2}T} \int_{x(T)=x(0)}{\cal D}x(\tau)\
e^{-\int_0^T d\tau\Bigl( \frac{1}{4}\dot{x}^{2} +ie\,\dot x\cdot A(x(\tau))+\sigma(x(\tau)) \Bigr)},
\end{eqnarray}
where coefficient $N$ before the integral reflects the $2N$ degrees of freedom of our scalar $O(N)$
model.\footnote{If we were working only with one neutral scalar field, this coefficient would be $\frac{1}{2}$.}
This path integral is to be calculated using the Gauss formula
\begin{equation}\label{gaussian}
\int dx\ e^{-x\cdot {\cal{O}}\cdot x+2b\cdot x}\sim {({\rm det} ({\cal{O}}))}^{-\frac{1}{2}}e^{b\cdot
{\cal{O}}^{-1}\cdot b},
\end{equation}
with ${\cal{O}}^{-1}$ the inverse of the operator ${\cal{O}}$. To build ${\cal{O}}^{-1}$, we have to be careful
about the zero eigenvalues of ${\cal{O}}$, which are to be excluded from its spectrum. To deal with these zero
modes,\footnote{In the eigenvalue equation ${\cal{O}}\psi=\lambda\psi$ these zero-modes correspond to constant
eigenfunctions $\psi=$const.} contained in the coordinate path integral $\int {\cal{D}}x$ as constant loops,
they are to be separated from their orthogonal non-zero modes $y(\tau)$ by evaluating $x(\tau)$ around the loop
center of mass $x_{0}$. In other words, replacing $x^{\mu}(\tau)$ by $x^{\mu}(\tau)=x^{\mu}_0+ y^{\mu}(\tau)$
with
$$\int_{0}^{T} d\tau \,y^{\mu}(\tau )=0,$$ the coordinate path integral reduces to an integral over the
relative coordinate $y(\tau)$
\begin{eqnarray}\label{split}
\int{\cal {D}}x=\int dx_0 \int {\cal {D}} y.
\end{eqnarray}
In this way, the effective action $\Gamma[\sigma]$ is expressed by an effective Lagrangian ${\cal{L}}_{eff}$,
representing as an integral over the space of all loop with fixed common center of mass $x_{0}$ \cite{worev},
\begin{eqnarray}
\Gamma[\sigma]=\int dx_{0}\ {\cal{L}}_{eff}[\sigma; x_{0}].
\end{eqnarray}
At this stage, we restrict the background to be constant. Using Fock-Schwinger gauge centered at $x_0$ we may
take $ A_{\mu}(x)$ in (\ref{rawGamma}) to be of the form \cite{focksch}
\begin{eqnarray}\label{fockschwinger}
A_{\mu}(x)={1\over 2}y^{\nu} F_{\nu\mu},
\end{eqnarray}
where $F_{\mu\nu}$ is the constant field-strength tensor of $A_{\mu}$. After removing the zero-modes, the
operator ${d^2\over {d\tau}^2} -2ie F {d\over d\tau}$ becomes invertible. The worldline Green's function is then
given by \cite{worev}
\begin{eqnarray*}
2\bigl\langle\tau_i| {\biggl( {d^2\over {d\tau}^2} -2ie F {d\over d\tau} \biggr) }^{-1}
|\tau_j\bigr\rangle\equiv {\cal G}_{B}(\tau_i,\tau_j),
\end{eqnarray*}
where
\begin{eqnarray}\label{S3-6}
{\cal G}_{B}(\tau_i,\tau_j)\equiv {T\over 2{({\cal Z})}^2}\biggl({{\cal Z}\over{{\rm sin}({\cal Z})}} {\rm
e}^{-i{\cal Z}\dot G_{Bij}} +i{\cal Z}\dot G_{Bij} -1\biggr),
\end{eqnarray}
with ${\cal{Z}}\equiv eFT$ and
\begin{eqnarray}\label{AA-3}
G_{Bij}&\equiv& G_B(\tau_i,\tau_j)\equiv | \tau_i-\tau_j| -{{(\tau_i-\tau_j)}^2\over T},\nonumber\\
\dot{G}_{Bij}&\equiv& \dot G_B(\tau_i,\tau_j)\equiv{\rm sign}(\tau_i-\tau_j)-2{{(\tau_i-\tau_j)}\over T}.
\end{eqnarray}
The Green's function (\ref{S3-6}) is used as the correlation function for the coordinate field
\begin{eqnarray}\label{wicky}
\langle y^{\mu}(\tau_i)y^{\nu}(\tau_j)\rangle= - g^{\mu\nu}{\cal G}_{B}(\tau_i,\tau_j).
\end{eqnarray}
To proceed, we also need to calculate the free worldline path integral
\begin{eqnarray}\label{aa-1}
\int {\cal D} y\, \exp\left(- \int_0^T d\tau \Bigl(\frac{1}{4}{\dot y}^2+ie\,\dot y\cdot A(y(\tau))\Bigr)
\right)\Bigg|_{A_{\mu}=\frac{1}{2}y^{\nu}F_{\mu\nu}}
&=&\mbox{Det}'^{-\frac{1}{2}}\bigg[-\frac{d^{2}}{d\tau^{2}}+2ieF\frac{d}{d\tau}\bigg]
\nonumber\\
&=&(4\pi T)^{-\frac{D}{2}}\
\mbox{Det}'^{-\frac{1}{2}}\bigg[{\mathbf{1}}-2ieF\left(\frac{d}{d\tau}\right)^{-1}\bigg]\nonumber\\
&=&(4\pi T)^{-{D\over 2}}{\rm det}^{-{1\over 2}} \biggl[{{\rm sin}(eFT)\over eFT}\biggr].
\end{eqnarray}
This result is obtained using the method introduced in \cite{worev}. On the first and second lines, the primes
denote the absence of the zero modes in the determinant.
\par
Let us now turn back to the effective action $\Gamma[\sigma]$ from (\ref{rawGamma}) containing the interaction
term $\sigma$. To find the $n$-point vertex function, we recall from quantum field theory that one-particle
irreducible (1PI) $n$-point function can be obtained from one-loop action $\Gamma[\sigma]$ by a $n$-fold
functional differentiation with respect to $\sigma$. In momentum space, this operation is implemented by
replacing the background to a sum of plane waves,
\begin{eqnarray}
\sigma(x)=\sum_{i=1}^n{\rm e}^{ip_i\cdot x},
\end{eqnarray}
and picking out the term containing every $p_{i}$ only once. We arrive at
\begin{eqnarray}
\Gamma_{\rm 1PI}[p_1,\ldots,p_n] &=&N(-1)^n
\int_{0}^{\infty}\frac{dT}{T}e^{-m^{2}T} \int_0^T \prod_{i=1}^n d\tau_i \nonumber\\
 &&
 \times
\int dx_0 \int {\cal D} y \, \exp \Bigl( i\sum_{i=1}^{n} p_i\cdot x_{i} \Bigr) e^{-\int_0^T d\tau\bigl(
\frac{1}{4}\dot{x}^{2}+ie\,\dot x\cdot A(x(\tau)) \bigr)}.
\end{eqnarray}
Having $x_i\equiv x(\tau_i)=x_0+y(\tau_i)$, the $x_0$-integral leads to energy-momentum conservation
\begin{eqnarray*}
\int d x_0\, {\rm \exp}\left({ix_0\cdot \sum_{i=1}^n p_i}\right) = {(2\pi)}^D\delta\Bigl(\sum_{i=1}^n p_i\Bigr).
\end{eqnarray*}
To perform the $y$ integration, we use the result from (\ref{aa-1}). This leads to the following parameter
integral
\begin{eqnarray}\label{scalarmaster}
\Gamma_{\rm 1PI}[p_1,\cdots,p_n] &=&N(-1)^n {(2\pi )}^D\delta \left(\sum\limits_{i=1}^{n}
p_i\right)\int_{0}^{\infty}\frac{dT}{T} {(4\pi T)}^{-{D\over 2}}\, e^{-m^2T}\,{\rm det}^{-{1\over 2}}
\biggl[{{\rm sin}(eFT)\over eFT}\biggr]
\nonumber\\
&&\times \prod_{i=1}^n \int_0^T d\tau_i \exp\left(\frac{1}{2}\sum_{i,j=1}^n{\cal G}_{B}(\tau_i,\tau_j)\ p_i\cdot
p_j\right).
\end{eqnarray}
To proceed, we will replace ${\cal G}_{B}(\tau_i,\tau_j)$ by $\bar{\cal G}_{B}(\tau_i,\tau_j)$, which is defined
by
\begin{eqnarray}
\bar{\cal G}_B(\tau_i,\tau_j) \equiv {\cal G}_B(\tau_i,\tau_j) - {\cal G}_B(\tau,\tau) = {T\over 2{\cal Z}},
\biggl( {e^{-i\dot G_{Bij}{\cal Z}}-\cos{\cal Z}\over\sin{\cal Z}} +i\dot G_{Bij}\biggr),
\end{eqnarray}
with the coincidence limit
\begin{eqnarray}\label{coincalG}
{\cal G}_{B}(\tau,\tau)&=& {T\over 2{{\cal Z}}^2} \biggl({\cal Z}\cot{\cal Z}-1 \biggr)
\nonumber\\
\dot {\cal G}_B(\tau,\tau) &=& i{\rm cot}{\cal Z} -{i\over {\cal Z}},
\end{eqnarray}
that are found by applying the relations
\begin{eqnarray}\label{coincidencerules}
\dot G_B(\tau,\tau)=0,\quad \dot G_B^2(\tau,\tau)=1.
\end{eqnarray}
After replacing $\tau_{i}\rightarrow T u_{i}$ in (\ref{AA-3}), the 1PI $n$-point vertex function in the presence
of a electromagnetic background field is given by
\begin{eqnarray}\label{masterwl}
\Gamma_{\rm 1PI}[p_1,\cdots,p_n] &=&N(-1)^n {(2\pi )}^D\delta \left(\sum\limits_{i=1}^{n}
p_i\right)\int_{0}^{\infty}\frac{dT}{T^{1-n}} {(4\pi T)}^{-{D\over 2}}\, e^{-m^2T}\,{\rm det}^{-{1\over 2}}
\biggl[{{\rm sin}(eFT)\over eFT}\biggr] \nonumber\\
&&\times \prod_{i=1}^n \int_0^1 du_i \exp\left(\sum_{i<j=1}^n\bar{\cal G}_B(u_i,u_j)\  p_i\cdot p_j\right).
\end{eqnarray}
At this stage, a constant magnetic field is chosen for the background. With the $B$-field chosen along the
z-axis, we introduce matrices $g_{\perp}$ and $g_{\parallel}$ projecting on $x,y$ and z, $\tau$ planes
respectively
\begin{equation}
\hat{F} = \left(
\begin{array}{*{4}{c}}
0&1&0&0\\
-1&0&0&0\\
0&0&0&0\\
0&0&0&0
\end{array}
\right),\ \  g_{\perp}\equiv \left(
\begin{array}{*{4}{c}}
1&0&0&0\\
0&1&0&0\\
0&0&0&0\\
0&0&0&0
\end{array}
\right),\ \  g_{\parallel}\equiv \left(
\begin{array}{*{4}{c}}
0&0&0&0\\
0&0&0&0\\
0&0&1&0\\
0&0&0&1
\end{array}
\right),\nonumber\\
\label{defBmatrices}
\end{equation}
with $\hat{F}\equiv \frac{F}{B}$. The determinant factor in (\ref{masterwl}) becomes
\begin{eqnarray}
{\rm det}^{-{1\over 2}} \biggl[{\sin{\cal Z}\over {{\cal Z}}} \biggr]={z\over{\sinh z}},
\end{eqnarray}
where we have introduced $z\equiv eBT$. Further, the worldline Green's function (\ref{S3-6}) is given by
\begin{eqnarray}\label{KK1}
\bar{\cal G}_{B}(u_i,u_j) =G_{Bij}\,{g_{\parallel}} -\frac{T}{2z}\left(\frac{\cosh(z\dot{G}_{Bij})}{\sinh
z}-\coth z\right) {g_{\perp}}+{T\over{2z}}\biggl({\sinh(z\dot G_{Bij})\over\sinh z} -\dot
G_{Bij}\biggr)i\hat{F}.
\end{eqnarray}
As was stated before, it is desirable to obtain the effective vertex function in the presence of strong magnetic
field. Let us therefore look at the worldline Green's function (\ref{KK1}) in the limit of strong magnetic
field. It is allowed to rearrange the parameters as $u_{1}>u_{2}>\cdots>u_{n-1}>u_n$. By this convention, it is
seen that $z\dot{G}_{Bij}<z$. We obtain
\begin{eqnarray} \label{proplimit}
\lim_{B \to \infty}\bar{\cal G}_{B}(u_i,u_j)=G_{Bij}\,{g_{\parallel}} +\frac{1}{2| eB
|}{g_{\perp}}-\frac{i}{2eB}\dot G_{Bij}\hat{F}.
\end{eqnarray}
Moreover, the determinant behaves in this limit as
\begin{eqnarray} \label{detlimit}
 \lim_{B \to \infty}\frac{z}{\sinh z}\approx 2ze^{-z}.
\end{eqnarray}
In (\ref{proplimit}) and (\ref{detlimit}), we choose the sign of $eB>0$ and we denote it by $|eB|$ whenever $eB$
arises from a term that is even in terms of it. Replacing (\ref{proplimit}) and (\ref{detlimit}) in
(\ref{masterwl}) for dimension $D=4$ yields\footnote{By $[\xi]\ p_i\cdot p_j$ for $\xi=g_{\parallel}$,
$g_{\perp}$ or $\hat{F}$ in (\ref{xy-1}), we mean $\xi^{\mu\nu} p_{i\mu} p_{j\nu}$. Here, $i$ and $j$ denote the
$i$-th and $j$-th incoming momentum; $\mu$ and $\nu$ denote the ${\mu}$-th and ${\nu}$-th  component of the four
vector.}
\begin{eqnarray}\label{xy-1}
\lefteqn{\hspace{-1.5cm}\Gamma_{\rm 1PI}[p_1,\cdots,p_n] = N| eB|{(-1)}^n (2\pi^{2})\ \delta
\left(\sum\limits_{i=1}^{n}
p_i\right)\int_{0}^{\infty}\frac{dT}{T^{2-n}} \, e^{-(m^2+| eB|)T}}\nonumber\\
&&\times \prod_{i=1}^n \int_0^1 du_i \exp\left(\sum_{i<j=1}^n[G_{Bij}\,{g_{\parallel}} +\frac{1}{2| eB
|}{g_{\perp}}-\frac{i}{2eB}\dot G_{Bij}\hat{F}]\ p_i\cdot p_j\right).
\end{eqnarray}
To make the meaning of this equation more transparent, each part will be considered independently. We refer to
the three terms in the exponent as $g_{\parallel}$, $g_{\perp}$ and $\hat{F}$ terms. Among them, $g_{\perp}$ and
$\hat{F}$ terms are independent of $T$, thus can be taken out of $T$-integral.
\par
For $g_{\perp}$ term we have
\begin{eqnarray}
\exp\left(\sum_{i<j=1}^n\frac{{g_{\perp}}}{2| eB |}\ p_i \cdot p_j\right)=\exp\left(\frac{1}{2| eB
|}({{\mathbf{p}}_1}_{\perp} \cdot {{\mathbf{p}}_2}_{\perp}+{{\mathbf{p}}_1}_{\perp} \cdot
{{\mathbf{p}}_3}_{\perp}+\cdots+{{\mathbf{p}}_{n-1}}_{\perp} \cdot {{\mathbf{p}}_{n}}_{\perp})\right),
\end{eqnarray}
where ${\mathbf{p}}_{\perp}\equiv (p_{x},p_{y})$. Applying the energy-momentum conservation
$\sum\limits_{i=1}^{n}{\mathbf{p}}_{i\perp}=0 $, we get
\begin{eqnarray}\label{g-perpterm}
\exp\left(\sum_{i<j=1}^n\frac{{g_{\perp}}}{2| eB |}\ ({\mathbf{p}}_i \cdot
{\mathbf{p}}_j)_{\perp}\right)=\exp\left(-\frac{1}{4| eB |}\sum_{i=1}^{n}{\mathbf{p}}^{2}_{i\perp}\right).
\end{eqnarray}
This is exactly the damping factor which appears also in \cite{EV}, where the effective action of NJL-model is
calculated in the LLL approximation.
\par
To calculate the $\hat{F}$ term, we use (\ref{AA-3}) with the replacement $\tau_{i}\to u_{i}T$, and are led to
\begin{eqnarray}\label{xxy-1}
\exp\left(-\frac{i}{2 eB }\sum_{i<j=1}^n[{\rm sign}(u_i - u_j) - 2 (u_i - u_j)]\hat{F}\ p_i \cdot p_j\right).
\end{eqnarray}
To build some parallels to the (modified) noncommutative effective field theory which should arise in the large
$B$-limit, we introduce the (noncommutativity) parameter
\begin{eqnarray}\label{theta}
\theta_{ab}\equiv \frac{1}{eB}\epsilon_{ab},
\end{eqnarray}
with $\epsilon_{ab}$ the ordinary antisymmetric tensor of rank two and indices $a$ and $b$ denoting transverse
directions, {\it i.e.} $x$ and $y$ coordinates. Using this new definition the product on the r.h.s. of
(\ref{xxy-1}) becomes
\begin{eqnarray}\label{crossproduct}
\frac{1}{eB}\hat{F}\ p_i \cdot p_j=p_{i}^{\ a}\theta_{ab}p_{j}^{\ b}=:{\mathbf{p}}_{i}\times {\mathbf{p}}_{j}.
\end{eqnarray}
In the definition of the cross product, we will skip the subscript $\perp$ for the transverse coordinates.
Adhering to our rearrangement for $u_{1}>u_{2}>\cdots>u_{n-1}>u_n$ and using the freedom to choose the zero
somewhere in the world loop for setting $u_n=0$, we have
\begin{eqnarray}
\sum_{i<j=1}^n[({\mathbf{p}}_i\times {\mathbf{p}}_j)(u_i - u_j)]=({\mathbf{p}}_1\times {\mathbf{p}}_2)(u_1 -
u_2)+({\mathbf{p}}_1\times {\mathbf{p}}_3)(u_1 - u_3)+\cdots+({\mathbf{p}}_1\times {\mathbf{p}}_n)(u_1)
\nonumber\\&& \hspace{-13cm}({\mathbf{p}}_2\times {\mathbf{p}}_3)(u_2 - u_3)+\cdots+({\mathbf{p}}_2\times
{\mathbf{p}}_n)(u_2)+\cdots+({\mathbf{p}}_{n-1}\times {\mathbf{p}}_n)(u_{n-1}).
\end{eqnarray}
Using further the energy-momentum conservation and the antisymmetry property of the cross-product, defined in
(\ref{crossproduct}), we obtain
\begin{eqnarray}
\sum_{i<j=1}^n[({\mathbf{p}}_i\times {\mathbf{p}}_j)(u_i - u_j)]=0.
\end{eqnarray}
The $\hat{F}$-term becomes therefore
\begin{eqnarray}\label{F-term}
\exp\left(-\frac{i}{2 eB }\sum_{i<j=1}^n\ [\dot G_{Bij}\hat{F}]\ p_i \cdot p_j\right)=
\exp\left(-\frac{i}{2}\sum\limits_{i<j=1}^n\ {\mathbf{p}}_i\times {\mathbf{p}}_j\right).
\end{eqnarray}
The same phase factor also appears in the Fourier transform of the vertices in the ordinary noncommutative field
theory \cite{DNS} that is defined by replacing the ordinary product of functions by the Moyal $\star$-product.
\par
Putting now (\ref{g-perpterm}) and (\ref{F-term}) together in (\ref{xy-1}), we finally obtain the 1PI $n$-point
function for the composite field $\sigma$ in a strong magnetic field
\begin{eqnarray}\label{finworld}
\lefteqn{\hspace{-2cm}\Gamma_{\rm 1PI}[p_1,\cdots,p_n] = N| eB |{(-1)}^n (2\pi^{2} )\ \delta
\left(\sum\limits_{i=1}^{n} p_i\right)\exp\left(-\frac{1}{4| eB
|}\sum\limits_{i=1}^n{\mathbf{p}}^{2}_{i\perp}\right) \exp\left(-\frac{i}{2}\sum\limits_{i<j=1}^n
{\mathbf{p}}_i\times {\mathbf{p}}_j\right)
}\nonumber\\
&&\times\int_{0}^{\infty}\frac{dT}{T^{2-n}} e^{-(m^2+| eB |)T}\,\prod_{i=1}^n \int_0^1 du_i
\exp\left(\sum_{i<j=1}^n[G_{Bij}\,{g_{\parallel}}]\ p_i\cdot p_j\right).
\end{eqnarray}
Comparison with the 1PI $n$-point vertex function
\begin{eqnarray}\label{turnoff}
\Gamma_{\rm 1PI}[p_1,\cdots,p_n] &=& N(-1)^n {(2\pi )}^D\delta \left(\sum\limits_{i=1}^{n}
p_i\right)\int_{0}^{\infty}\frac{dT}{T^{1-n}} {(4\pi T)}^{-{D\over
2}}\, e^{-m^2T}\, \nonumber\\
&&\times \prod_{i=1}^n \int_0^1 du_i \exp\left(\sum_{i<j=1}^n G_{Bij}\ p_i\cdot p_j\right),
\end{eqnarray}
which is obtained by turning off the electromagnetic field in (\ref{masterwl}), reveals that the parallel sector
of (\ref{finworld}), involving the $\tau$ and $z$ directions, is up to some factor the same as the $\Gamma_{\rm
1PI}$ (\ref{turnoff}) in $D=2$ dimensions, provided that $m^2\rightarrow m^2+|eB|$. In the next section, using
the method introduced in \cite{EV}, we will find the $n$-point contribution to the effective action of the
composites $\sigma$ in the presence of a strong magnetic field in an appropriate LLL approximation. The result
will have common features with (\ref{finworld}). In particular, the phases (\ref{g-perpterm}) and (\ref{F-term})
reappear in the final result.
\subsubsection*{4\hspace{0.3cm}Effective Action of the Composite Field in the LLL Approximation}
\setcounter{section}{4} \setcounter{equation}{0} In the first part of this section, starting from the full
bosonic Green's function derived in the seminal work of J. Schwinger \cite{Schwinger} in the framework of
Schwinger proper time formalism, we will determine the propagator of a multidimensional complex scalar field
$\Phi$ in the LLL approximation. We then use the LLL propagator to determine the effective action of $n$
composite fields $\sigma$ in the LLL regime.
\par
Let us start with the Schwinger propagator
\begin{nedalph}\label{schwinger1a}
G(x',x'')=P(x',x'')D(x'-x''),
\end{eqnarray}
with
\begin{eqnarray}\label{schwinger1b}
P(x',x'')\equiv \exp\left(ie \int_{x''}^{x'}d{\xi^\mu}A_\mu(\xi)\right),
\end{eqnarray}
and
\begin{eqnarray}\label{schwinger1c}
D(x'-x'')\equiv \frac{-1}{(4\pi)^2}\int\limits_{0}^{\infty}\frac{ds}{s^2} e^{-ism^2}\exp\left(\frac{-1}{2}{\rm
tr}\ln\bigg[\frac{\sinh eFs}{eFs}\bigg]\right)
\exp\bigg[\frac{i}{4}(x'-x'')eF\coth(eFs)(x'-x'')\bigg].\nonumber\\
\end{nedalph}

\vspace{-1cm}\noindent In the symmetric gauge
\begin{eqnarray*}\label{symmetricgauge}
A_\mu=\frac{B}{2}\left(0,x_{2},-x_{1},0\right),
\end{eqnarray*}
the part consisting of the Schwinger line integral is equal to
\begin{eqnarray}\label{schwingerline}
P(x',x'')=e^{\frac{ieB}{2}\epsilon^{ab}x'_{a}x''_{b}},\qquad\qquad a,b=1,2.
\end{eqnarray}
Here, as in the previous section, $B$ is chosen to be a constant magnetic background field in $x_{3}$-direction.
In the translationally invariant part $D(x'-x'')$ in (\ref{schwinger1c}), it is easy to show that
\begin{eqnarray}\label{Q1}
x^{\alpha}{[eF\coth(eFs)]_{\alpha}}^{\beta}x_\beta=\frac{1}{s}{\mathbf{x}}^2_\parallel- eB
\coth(eBs){\mathbf{x}}^2_\perp,
\end{eqnarray}
with ${\mathbf{x}}^2_\parallel\equiv x^2_0-x^2_3$ and ${\mathbf{x}}^2_\perp\equiv x^2_1+x^2_2$, and
\begin{eqnarray}\label{Q2}
\exp\left(-\frac{1}{2}{\rm tr}\ln\bigg[\frac{\sinh (eFs)}{eFs}\bigg]\right)=\frac{eBs}{\sin (eBs)}.
\end{eqnarray}
Putting (\ref{Q1}) and (\ref{Q2}) in (\ref{schwinger1c}) and  after taking the Fourier transformation, we obtain
\begin{eqnarray}
\tilde{D}(k)=-\int_{0}^{\infty}\frac{ds}{\cosh \left(eBs\right)}\exp\left(-s\left(m^2-k_0^2+{\bf k}_{\perp}^2
{\tanh (eBs)\over eBs}+k_3^2 \right)\right).
\end{eqnarray}
Here, we assume again that $eB>0$ and replace $eBs\to s$. We arrive at\footnote{As it was stated earlier $eB$ is
chosen to be $| eB |$ whenever it appears in even terms.}
\begin{eqnarray}\label{transinv}
\tilde{D}(k)=-\frac{1}{eB}\int_{0}^{\infty}\frac{ds}{\cosh s}\ e^{-s\rho-\alpha\tanh(s)},
\end{eqnarray}
with $\alpha\equiv\frac{{\bf k}_{\perp}^2}{| eB |}$ and $\rho\equiv\frac{m^2-{\bf k}_{\parallel}^2}{| eB |}$.
Further, ${\bf k}_{\parallel}^2\equiv k_0^2-k_3^2$ and ${\bf k}_{\perp}^2\equiv k_1^2+k_2^2$ are introduced. To
determine the regime of LLL dominance of $\tilde{D}(k)$, we use first (\ref{Lagexp}) from appendix B, to
obtain\footnote{A similar method was also used in \cite{chodos}.}
\begin{eqnarray}\label{DK}
\tilde{D}(k)=\frac{e^{-\alpha}}{| eB |}\sum_{n'=0}^\infty (-1)^{n'}L_{n'}^{(-1)} (2\alpha)\int_{0}^{\infty}ds\
\frac{e^{-s(\rho+2n')}}{\cosh s},
\end{eqnarray}
where $L_{n'}^{(\beta)}$ is the generalized Laguerre polynomial. Then, using (\ref{intcos}) from appendix C, we
get
\begin{eqnarray}\label{Dpsi}
\tilde{D}(k)=\frac{e^{-\alpha}}{2| eB |}\sum_{n'=0}^\infty (-1)^{n'} L_{n'}^{(-1)}
(2\alpha)\Bigg[\psi\left(\frac{\rho+2n'+1}{4}\right)-\psi\left(\frac{\rho+2n'+3}{4}\right)\Bigg].
\end{eqnarray}
Next, we expand digamma function $\psi(z)$ \cite{Handbook} according to
\begin{eqnarray}\label{psiexp}
\psi(1+z)=-\gamma+\sum_{m'=1}^\infty\frac{z}{m'(m'+z)},
\end{eqnarray}
and arrive at
\begin{eqnarray}\label{propexpansion}
\tilde{D}(k)=\frac{e^{-\alpha}}{2| eB |}\sum_{n'=0,m'=1}^\infty (-1)^{n'}L_{n'}^{(-1)}
(2\alpha)\Bigg[\frac{\rho+2n'-3}{m'(4m'+\rho+2n'-3)}-\frac{\rho+2n'-1}{m'(4m'+\rho+2n'-1)}\Bigg].
\end{eqnarray}
Before we continue and determine the LLL form of $\tilde{D}(k)$, we note that when the dynamics of a particle is
stationary, as in the present case with pure constant magnetic background, the energy spectrum can be read from
the poles of the propagator. In other words, the energy spectrum of a particle in magnetic background that is
obtained from the relativistic Klein-Gordon equation, coincides with the poles of the propagator that entails
corrections due to the background. This fact enables us to obtain the  effective propagator in the LLL dominant
regime.
\par
The energy spectrum of a scalar field in the presence of magnetic background is known to be (see {\it{e.g.}}
\cite{Zuber})
\begin{eqnarray}\label{energyspectrum}
E_{\ell'}(k)=\sqrt{m^2+| eB |(2\ell'+1)+k_3^2} ,\hspace{1cm}\mbox{for}\hspace{1cm}\ell'=0,1,2,\cdots\infty.
\end{eqnarray}
Choosing $\ell'=0$ the energy of the lowest Landau level (LLL) is given by
\begin{eqnarray}\label{LLLenergy}
E_{0}^{\ 2}=k_3^{\ 2}+m^2+|eB|.
\end{eqnarray}
To find the LLL form of $\tilde{D}(k)$, we determine the variables $m$ and $n$ in (\ref{propexpansion}) such
that the LLL energy arising from the poles of the propagator (\ref{propexpansion}) coincides with
(\ref{LLLenergy}). Using the notation introduced before, we get $E_{0}^2-k_3^2-m^2={\bf k}_{\parallel}^2-m^2=|
eB |$ that yields $\rho=-1$. As it turns out, the only valid choice satisfying the first pole equation
$m'(4m'+2n'-4)=0$ from the first denominator in (\ref{propexpansion}) is $(m'=1, n'=0)$. The second pole
equation $m'(4m'+2n'-2)=0$ from the second denominator in (\ref{propexpansion}) does not have any valid solution
and therefore does not contribute to the LLL propagator. Thus, plugging $(m'=1,n'=0)$ in (\ref{propexpansion})
and using $L_{0}^{(-1)}(2\alpha)=1$, we get
\begin{eqnarray}\label{LLLD}
{\tilde{D}_{LLL}(k)}=e^{-\frac{{\bf k}_{\perp}^2}{|eB|}} \frac{2}{{\bf k}_{\parallel}^2-(m^2+ |eB|)}.
\end{eqnarray}
In the coordinate space the LLL effective propagator (\ref{schwinger1a}-c) can therefore be written in the form
\begin{eqnarray}
G_{LLL}(x',x'')=P(x',x''){\cal F}^{-1} \Biggl\{ \tilde{D}_{LLL}(k) \Biggr\},
\end{eqnarray}
where $P(x',x'')$ is defined in (\ref{schwingerline}) and ${\cal{F}}^{-1}$ denotes the inverse Fourier
transform. From now on, we drop the subscript LLL, but it is always assumed. $G(x',x'')$ obviously factorizes
into two independent transverse and longitudinal parts
\begin{nedalph}\label{sepa}
G(x',x'')=G_\perp({\mathbf{x}}'_\perp,{\mathbf{x}}''_\perp)
G_\parallel({\mathbf{x}}'_\parallel-{\mathbf{x}}''_\parallel),
\end{eqnarray}
where the transverse part is
\begin{eqnarray}\label{sepb}
G_\perp({\mathbf{x}}'_\perp,{\mathbf{x}}''_\perp)=\frac{| eB |}{2\pi}e^{\frac{ieB}{2}\epsilon^{ab}
x'_{a}x''_{b}}\ e^{-\frac{| eB |}{4}({\mathbf{x}}'_\perp-{\mathbf{x}}''_\perp)^2},
\end{eqnarray}
including the Schwinger line integral (\ref{schwingerline}) and the Fourier transform of the phase factor
$e^{-{\mathbf{k}}_{\perp}^{2}/|eB|}$ from (\ref{LLLD}), and the longitudinal part is
\begin{eqnarray}\label{sepc}
 G_\parallel({\mathbf{x}}'_\parallel-{\mathbf{x}}''_\parallel)={\cal F}^{-1} \Biggl\{
 \frac{1}{{\bf k}_{\parallel}^2-(m^2+
| eB |)}\Biggr\}.
\end{nedalph}

\vspace{-0.5cm}\noindent At this stage, we have all the necessary tools to calculate the $n$-point vertex
function $\Gamma_{n\sigma}$ for $n$-composite fields $\sigma$. It is obtained through
\begin{eqnarray}
\Gamma_{n\sigma}=\int d^{4}x_{1}\cdots d^{4}x_{n} \Bigl[ \sigma(x_1)  G (x_1,x_2)\sigma(x_2)G (x_2,x_3) \cdots
\sigma(x_n)G(x_n,x_1)\Bigr].
\end{eqnarray}
Using now (\ref{sepa}-c) for $G(x',x'')$, inserting the Fourier transform of the composite fields $\sigma$ and
carrying out the integrations over ${\mathbf{x}}$, the $n$-point contribution to the effective action reads
\begin{eqnarray}\label{finprop}
\Gamma_{n\sigma}&=&2\pi N| eB | \int d^{2}x_{1\parallel}\cdots d^{2}x_{n\parallel}\  \frac{d^4
p_1}{(2\pi)^4}\cdots \frac{d^4 p_n}{(2\pi)^4} \delta^2\left(\sum\limits_{i=1}^{n}
{\mathbf{p}}_{i\perp}\right)\exp \left(i\sum\limits_{i=1}^{n}
{\mathbf{p}}_{i\parallel}\cdot {\mathbf{x}}_{i\parallel} \right)\nonumber\\
&&\times \exp\left( -\frac{1}{4| eB |}\sum\limits_{i=1}^{n} {\mathbf{p}}_{i\perp}^2 \right) \exp\left(
-\frac{i}{2}\sum_{i<j=1}^{n} {\mathbf{p}}_{i}\times
{\mathbf{p}}_{j}  \right)\nonumber\\
&&\times \Bigl[ \sigma(p_1)  G_\parallel
({\mathbf{x}}_{1\parallel},{\mathbf{x}}_{2\parallel})\sigma(p_2)G_\parallel({\mathbf{x}}_{2\parallel},{\mathbf{x}}_{3\parallel})
\cdots \sigma(p_n)G_\parallel({\mathbf{x}}_{n\parallel},{\mathbf{x}}_{1\parallel})\Bigr].
\end{eqnarray}
The cross product on the second line is defined in (\ref{crossproduct}) and includes only the transverse
coordinates of $p_{i}$ {\it i.e.} ${\mathbf{p}}_{i\perp}=(p_{i1},p_{i2})$. In the next section we compare this
result with the 1PI $n$-point vertex function (\ref{finworld}), which we obtained in the previous section in the
limit of strong magnetic field.
\subsubsection*{5\hspace{0.3cm}Modified Noncommutative Field Theory (NCFT)}
\setcounter{section}{5}\setcounter{equation}{0} Let us consider the $n$-point vertex function  (\ref{finworld})
which was found in the framework of the worldline formalism and the $n$-point effective action (\ref{finprop})
which was obtained in the LLL approximation. As for the parallel sector of the effective theory, similar to
(\ref{finworld}), (\ref{finprop}) shows a free propagation of the composite field $\sigma$ in the longitudinal
coordinates, with the effective square mass $m^2+|eB|$. The same dimensional reduction from $D=4$ to $D=2$
dimensions was also observed in the dynamics of fermion pairing in a constant magnetic field for an effective
NJL model \cite{Gusynin}.
\par
As for the transverse part, the general structure of both results (\ref{finworld}) and (\ref{finprop}) can be
compared with the general structure of an $n$-point vertex of a conventional NCFT (see for instance in
\cite{DNS})
\begin{eqnarray}\label{ncftvertex}
\int d^{D}x\ \overbrace{\phi(x)\star\cdots\star\phi(x)}^{\mbox{\small {n-times}}}=\int \frac{d^D
p_1}{(2\pi)^D}\cdots\frac{d^D p_n}{(2\pi)^D}\ \delta^D\left(\sum_{i=1}^{n} p_i\right) \exp\left(-\frac{i}{2}
\sum\limits_{i<j=1}^{n} {\mathbf{p}}_i \times {\mathbf{p}}_j\right) \phi(p_1)\cdots\phi(p_n) .\nonumber\\
\end{eqnarray}
Here, the ordinary Moyal $\star$-product is defined by
\begin{eqnarray}\label{star}
\phi(x)\star\phi(x)=e^{\frac{i}{2}{\theta}^{ij}\frac{\partial}{\partial y^{a}}\frac{\partial}{\partial z^{b}}}
\sigma(y)\sigma(z)\Bigg|_{y=z=x},\qquad a,b=1,2,
\end{eqnarray}
which reflects the noncommutativity in the $x_{1}$ and $x_{2}$ coordinates
\begin{eqnarray}\label{noncom}
\big[x_{a},x_{b}\big]=i\theta_{ab},\qquad a,b=1,2,
\end{eqnarray}
with $\theta_{ab}\equiv \theta\epsilon_{ab}$ and $\epsilon_{ab}$ the ordinary antisymmetric tensor of rank two.
Similarly, as was originally shown in \cite{EV} for a fermionic NJL model, the noncommutative feature of
(\ref{finworld}) and  (\ref{finprop}) manifests itself in the phase factor containing the cross product
${\mathbf{p}}_{i}\times {\mathbf{p}}_{j}=p_{i}^a\theta_{ab}p_{j}^b $ with $a,b=1,2$ with $\theta$ defined in
(\ref{theta}). However, here in contrast to the ordinary noncommutative field theory, in (\ref{finworld}) as
well as in (\ref{finprop}) an additional phase factor
\begin{eqnarray}\label{phasefactor}
\exp\left( -\frac{1}{4| eB |}\sum\limits_{i=1}^{n} {\mathbf{p}}_{i\perp}^2 \right),
\end{eqnarray}
appears which modifies the noncommutativity between the longitudinal coordinates (\ref{noncom}) to
\begin{eqnarray}
[x^a,x^b]=i\widehat{\theta}^{ab},\hspace{2cm} a,b=1,2.
\end{eqnarray}
The modified noncommutative parameter $\widehat{\theta}$ is given by \cite{EV}
\begin{eqnarray}\label{widetheta}
\widehat{\theta} = \frac{1}{|eB|}\left(
\begin{array}{*{4}{c}}
i&{\rm sign}(eB)\\
-{\rm sign }(eB)&i
\end{array}
\right).
\end{eqnarray}
Using this definition, the full phase factor which manifests the noncommutative properties of the effective
$n$-point vertex function can be rewritten as
$$
\exp\left(-\frac{1}{4|eB|}\sum\limits_{i=1}^{n} {\mathbf{p}}_{i\perp}^2
-\frac{i}{2}\sum_{i<j=1}^{n}{\mathbf{p}}_{i} \times {\mathbf{p}}_{j}\right)=\exp\left(
-\frac{i}{2}\sum_{i<j=1}^{n} {\mathbf{p}}_{i}\widehat{\times} {\mathbf{p}}_{j}\right), $$ with $
{\mathbf{p}}_{i}\widehat{\times} {\mathbf{p}}_{j}= p_{i}^a {\widehat\theta}_{ab}p_{j}^b, a,b=1,2$. In the
special case where the composites are independent of the longitudinal coordinates, ${\mathbf{x}}_{\parallel}$,
the effective action of $n$ composites, (\ref{finprop}), in coordinate space is
\begin{eqnarray}\label{effact}
\Gamma_{n\sigma}\sim \int d^{2}x_{\parallel}\ d^{2}x_{\perp}\ \underbrace{\sigma({\mathbf{x}}_{\perp})\
\widehat{\star} \cdots\widehat{\star}\ \sigma({\mathbf{x}}_{\perp})}_{\mbox{\small{n-times}}},
\end{eqnarray}
with the modified Moyal $\widehat{\star}$-product, defined by
\begin{eqnarray}\label{modstar}
\sigma({\mathbf{x}}_{\perp})\ \widehat{\star}\
\sigma({\mathbf{x}}_{\perp})=e^{\frac{i}{2}\widehat{\theta}^{ab}\frac{\partial}{\partial
y^{a}}\frac{\partial}{\partial z^{b}}} \sigma(y)\sigma(z)\Bigg|_{y=z=x},\qquad a,b=1,2.
\end{eqnarray}
Alternatively, the above phase factor (\ref{phasefactor}) can be absorbed in the definition of the composite
field $\sigma$, leading to a new {\it smeared} field \cite{EV}
\begin{eqnarray}
\Sigma(x)\equiv e^{\frac{{\vec{\nabla}}_{\perp}^{2}}{4|eB|}}\sigma(x)
\end{eqnarray}
In terms of the smeared fields $\Sigma$, the effective action $\Gamma_{n\Sigma}$ of $n$ composites in coordinate
space is similar to (\ref{effact}) with $\sigma$ replaced by $\Sigma$ and modified $\widehat{\star}$-product
replaced by ordinary $\star$-product (\ref{star}).
\par
It is worth to mention, that the damping phase factor (\ref{phasefactor}) protects the modified NCFT from the
appearance of the UV/IR mixing \cite{MRS} which appears in the ordinary NCFT. This can be shown by writing the
one-loop correction to the tree level propagator $G_{\sigma}^{tree}(p)$. The one-loop vertices can be read {\it
e.g.} from the (\ref{finprop}) for $n=4$. The explicit calculation, similar to what is performed in \cite{EV}
for NJL model, shows that when the loop momentum $\ell\rightarrow\infty$, the loop integral is convergent for
all external legs momenta $p$, even for ${\mathbf{p}}_{\perp}\rightarrow 0$.
\subsubsection*{6\hspace{0.3cm}Conclusion}
In this paper, using two apparently different methods, the effective action of $n$ composite scalar fields is
derived in the presence of a strong background magnetic field. In section 3, the 1PI $n$-point amplitude of
these composites is determined in the framework of worldline formalism. Following the standard manipulations, we
arrived first at the path integral representation of the $n$-point vertex function in the presence of a constant
magnetic field. As a novelty, we then took the limit of strong magnetic field and ended up with an effective
$n$-point vertex function which is similar to the vertex function of a  ``modified'' noncommutative field
theory. In comparison to the standard noncommutative field theory a phase factor occurs which protects the
effective field theory from inconsistencies due to the appearance of UV/IR mixing.

In section 4, we followed the method presented in \cite{EV} and calculated the contribution of $n$ composites to
the effective action in the regime of lowest Landau level (LLL) dominance. First, starting from the full
Schwinger propagator of the scalar field in the presence of a background field, we derived the effective
propagator of the theory in an appropriate LLL approximation. Then, using this effective Green's function, we
determined the $n$-point effective action in this approximation.

In section 5, we compared both results from section 3 and 4. We showed that here, as in NJL model in the
presence of constant magnetic background \cite{Gusynin}, a dimensional reduction from $D=4$ to $D=2$ occurs in
the longitudinal section of the effective theory where a free propagation of the composites are observed. In the
transverse section however, a modified noncommutative field theory arises where no UV/IR mixing occurs. The
emergence of a modified noncommutativity is due to a breakdown of the translational invariance which, in section
4, exhibits itself in an explicit dependence of the full Green function of the theory on a Schwinger phase line
integral. Although the Schwinger line integral does not appear explicitly in the worldline formalism in section
3, but in the heart of this formalism is in fact, the expectation value of the Wilson loop which is the
Schwinger phase line integral taken along a closed loop.

\subsubsection*{7\hspace{0.3cm}Acknowledgment}
The authors thank F. Ardalan for useful discussions.

\begin{appendix}
\subsubsection*{Appendix A}
\setcounter{section}{1} \setcounter{equation}{0} In this appendix, we show that in four dimension and in the
presence of a constant magnetic background field, dynamical symmetry breaking occurs and a condensate always
develops. This authenticate our assumption for introduction of the composite field $\sigma$ in section 2. The
condensate is expressed through the complex boson propagator $G(x',x'')\equiv i\langle
0|\phi^{\ast}(x')\phi(x'')|0 \rangle$
\begin{eqnarray} \label{condens1}
\langle 0|\phi^{\ast}\phi|0\rangle=-i\lim\limits_{x'\to
x''}G(x',x'')=-\frac{i|eB|}{(4\pi)^2}\int_{0}^{\infty}\frac{ds}{s\sin (|eB|s)} e^{-ism^2}.
\end{eqnarray}
To deal with the infinity in the integral arising from $s\to 0$ limit, we can either restrict the lower limit of
$s\rightarrow\frac{1}{\Lambda^2}$ or alternatively render it finite through $\zeta$-function regularization.
Here we opt for the latter. After rotating the contour of the $s$-integration by $s\rightarrow -is$, we
introduce $\mu$ in the exponent of  (\ref{condens1}) to get
\begin{eqnarray}\label{condens2}
\langle0|\phi^{\ast}\phi|0\rangle =-\frac{|eB|}{(4\pi)^2}\int_{0}^{\infty}ds\frac{s^{\mu-1}}{\sinh
(s)}e^{-s\frac{m^2}{|eB|}}.
\end{eqnarray}
In the spirit of dimensional regularization, we will analytically continue to $\mu\rightarrow0$. By making use
of the integral representation of the Hurwitz $\zeta$-function,
\begin{eqnarray}
\zeta(r,\nu)=\frac{1}{\Gamma(t)}\int_{0}^{\infty}dt\ t^{r-1}\frac{e^{-t\nu}}{1-e^{-t}},\qquad\mbox{Re }t>1,\
\mbox{Re}\nu>0.
\end{eqnarray}
Thus, the integration in (\ref{condens2}) can be written as
\begin{eqnarray}\label{condens3}
\langle0|\phi^{\ast}\phi|0\rangle
=\frac{-|eB|2^{1-\mu}}{(4\pi)^2}\Gamma(\mu)\zeta\left(\mu,\frac{m^2}{2|eB|}+\frac{1}{2}\right).
\end{eqnarray}
Expansion around the poles yields
\begin{eqnarray}\label{condens4}
\langle0|\phi^{\ast}\phi|0\rangle =-\frac{|eB|}{8\pi^2}\left( \frac{1}{\mu}-\gamma+{\cal{O}}(\mu) \right)\left(
\frac{m^2}{2|eB|}+\bigg[\ln\Gamma\left(\frac{m^2}{2|eB|}+\frac{1}{2}\right)-\frac{1}{2}\ln(2\pi)\bigg]\mu\right),
\end{eqnarray}
where we have used the identities
\begin{eqnarray}
\zeta(0,\nu)&=&\frac{1}{2}-\nu \nonumber\\
\frac{\partial}{\partial r}\zeta(r,\nu)|_{r=0}&=&\ln\Gamma(\nu)-\frac{1}{2}\ln(2\pi).
\end{eqnarray}
Keeping the finite terms in (\ref{condens4}) and taking the limit $m^2\rightarrow0$, we obtain
\begin{eqnarray}\label{condens5}
\langle0|\phi^{\ast}\phi|0\rangle &=&\frac{|eB|}{(4\pi)^2}\ln(2).
\end{eqnarray}
This relation shows that in the presence of a constant magnetic background field the expectation value of the
condensate is nonzero and the composite field $\sigma$ is always formed. As it was implied earlier the
condensate is proportional to $|eB|$.
\subsubsection*{Appendix B}
\setcounter{section}{2} \setcounter{equation}{0} In this appendix, we derive the useful formulae which will help
us to determine the LLL effective propagator in section 4. We start with the identity
\begin{eqnarray}
(1-z)^{(-\beta-1)}\exp\Bigl(\frac{\lambda z}{z-1}
\Bigr)=\sum\limits_{n'=0}^{\infty}L_{n'}^{(\beta)}(\lambda)z^{n'} ,\qquad\mbox{for}\qquad |z|<1,
\end{eqnarray}
where $L_{n'}^{(\beta)}$ are the generalized Laguerre polynomials \cite{Handbook}. For $\beta=-1$, it can be
written as
\begin{eqnarray}\label{B2}
\exp\Bigg[\frac{\lambda}{2}\Bigl(\frac{z+1}{z-1}\Bigr)\Bigg]=e^{\frac{-\lambda}{2}} \sum_{n'=0}^\infty
L_{n'}^{(-1)}(\lambda)z^{n'},\ \ \ \qquad\mbox{for}\qquad |z|<1,
\end{eqnarray}
with $L_{n'}^{(\beta)}$ satisfying
\begin{eqnarray}
L_{n'}^{(-1)}(\lambda)= L_{n'}(\lambda)-L_{n'-1}(\lambda).
\end{eqnarray}
Defining $z=-e^{-2s}$ and using (\ref{B2}), we obtain\footnote{The condition $|z|<1$ is satisfied because in
(\ref{DK}) $s=0$ is a singular point. To deal with this infinity the integral is always proper-time regularized
by having by replacing the integration interval from $[0,+\infty]$ to $[\frac{1}{\Lambda^2},\infty]$.}
\begin{eqnarray}\label{Lagexp}
e^{- \alpha \tanh (s)}= \exp\Bigg[- \alpha \left(\frac{e^s -e^{-s}}{e^s +e^{-s}}\right)\Bigg]=e^{-\alpha}
\sum_{n'=0}^\infty(-1)^{n'} L_{n'}^{(-1)}(2\alpha)e^{-2n's}.
\end{eqnarray}
\subsubsection*{Appendix C}
\setcounter{section}{3} \setcounter{equation}{0} In order to perform the integrals of the form
\begin{eqnarray}
\int_{0}^{\infty}dt  \frac{e^{-zt}}{\cosh(t)},
\end{eqnarray}
we start from the definition of digamma function \cite{Handbook}
\begin{eqnarray}
\psi(z)+\gamma=\int_{0}^{\infty}dt \Bigg[ \frac{e^{-t}-e^{-zt}}{1-e^{-t}}\Bigg]=\int_{0}^{\infty}dt \Bigg[
\frac{e^{-\frac{t}{2}}-e^{-(z-\frac{1}{2})t}}{\left(e^{\frac{t}{4}}-e^{-\frac{t}{4}}\right)\left(e^{\frac{t}{4}}+e^{-\frac{t}{4}}\right)}\Bigg].
\end{eqnarray}
Then it is easy to show that
\begin{eqnarray}
\psi\left(z+\frac{3}{4}\right)-\psi\left(z+\frac{1}{4}\right)=\int_{0}^{\infty}dt \Bigg[
\frac{e^{-zt}}{\left(e^{\frac{t}{4}}+e^{-\frac{t}{4}}\right)}\Bigg].
\end{eqnarray}
Replacing $ t\rightarrow 4t$, we finally arrive at
\begin{eqnarray}\label{intcos}
\int_{0}^{\infty}dt  \frac{e^{-zt}}{\cosh(t)}=\frac{1}{2} \Bigg[
\psi\left(\frac{z+3}{4}\right)-\psi\left(\frac{z+1}{4}\right)\Bigg].
\end{eqnarray}
\end{appendix}


\begin{thebibliography}{99}
\bibitem{nc}
A. Connes, {\it Noncommutative Geometry}, Academic Press (1994).
\par
  A.~Connes, M.~R.~Douglas and A.~Schwarz,
  \textit{Noncommutative geometry and matrix theory: Compactification on tori},
  JHEP {\bf 9802}, 003 (1998)
  [arXiv:hep-th/9711162].
\bibitem{witten}
M  M.~R.~Douglas and C.~M.~Hull,
  \textit{D-branes and the noncommutative torus},
  JHEP {\bf 9802}, 008 (1998)
  [arXiv:hep-th/9711165].
\par
  F.~Ardalan, H.~Arfaei and M.~M.~Sheikh-Jabbari,
  \textit{Mixed branes and M(atrix) theory on noncommutative torus},
  [arXiv:hep-th/9803067].
\par
  F.~Ardalan, H.~Arfaei and M.~M.~Sheikh-Jabbari,
  \textit{Noncommutative geometry from strings and branes},
  JHEP {\bf 9902}, 016 (1999)
  [arXiv:hep-th/9810072].
\par
  F.~Ardalan, H.~Arfaei and M.~M.~Sheikh-Jabbari,
  \textit{Dirac quantization of open strings and noncommutativity in branes},
  Nucl.\ Phys.\ B {\bf 576}, 578 (2000)
  [arXiv:hep-th/9906161].
\par
C.-S. Chu, and P.-M. Ho, {\it Noncommutative open string and D-brane}, Nucl. Phys. {\bf B550} 151 (1999); {\it
Constrained quantization of open string in background B field and noncommutative D-brane}, Nucl. Phys. {\bf
B568} 447 (2000).
\par
  V.~Schomerus,
  \textit{D-branes and deformation quantization},
  JHEP {\bf 9906}, 030 (1999)
  [arXiv:hep-th/9903205].
\par
  N.~Seiberg and E.~Witten,
  \textit{String theory and noncommutative geometry},
  JHEP {\bf 9909}, 032 (1999)
  [arXiv:hep-th/9908142].
\bibitem{ym}
  T.~Filk,
  \textit{Divergencies in a field theory on quantum space},
  Phys.\ Lett.\ B {\bf 376}, 53 (1996).
\par
  C.~P.~Martin and D.~Sanchez-Ruiz,
  \textit{The one-loop UV divergent structure of U(1) Yang-Mills theory on
  noncommutative R**4},
  Phys.\ Rev.\ Lett.\  {\bf 83}, 476 (1999)
  [arXiv:hep-th/9903077].
\par
  A.~Armoni,
  \textit{Comments on perturbative dynamics of noncommutative Yang-Mills theory},
  Nucl.\ Phys.\ B {\bf 593}, 229 (2001)
  [arXiv:hep-th/0005208].
\par
  I.~Chepelev and R.~Roiban,
  \textit{Renormalization of quantum field theories on noncommutative R**d.  I: Scalars},
  JHEP {\bf 0005}, 037 (2000)
  [arXiv:hep-th/9911098].
\par
I. Ya. Aref'eva, D. M. Belov, and  A. S. Koshelev, {\it Two loop diagrams in noncommutative $\varphi^{4}$
theory},  Phys. Lett. {\bf B476} 431 (2000), [arXiv:hep-th/9912075].
\par
  J.~Gomis, K.~Landsteiner and E.~Lopez,
  \textit{Non-relativistic noncommutative field theory and UV/IR mixing},
  Phys.\ Rev.\ D {\bf 62}, 105006 (2000)
  [arXiv:hep-th/0004115].
\par
  C.~P.~Martin and F.~Ruiz Ruiz,
  \textit{Paramagnetic dominance, the sign of the $\beta$ function and UV/IR mixing in noncommutative U(1)},
  Nucl.\ Phys.\ B {\bf 597}, 197 (2001)
  [arXiv:hep-th/0007131].
\par
  A.~Matusis, L.~Susskind and N.~Toumbas,
  \textit{The IR/UV connection in the noncommutative gauge theories},
  JHEP {\bf 0012}, 002 (2000)
  [arXiv:hep-th/0002075].
\par
  F.~Ardalan and N.~Sadooghi,
  \textit{Axial anomaly in noncommutative QED on R**4},
  Int.\ J.\ Mod.\ Phys.\ A {\bf 16}, 3151 (2001)
  [arXiv:hep-th/0002143].
\par
  F.~Ardalan and N.~Sadooghi,
  \textit{Anomaly and nonplanar diagrams in noncommutative gauge theories},
  Int.\ J.\ Mod.\ Phys.\ A {\bf 17}, 123 (2002)
  [arXiv:hep-th/0009233].
\par
  F.~Ardalan and N.~Sadooghi,
  \textit{Planar and nonplanar Konishi anomalies and exact Wilsonian effective
  superpotential for noncommutative N = 1 supersymmetric U(1)},
  Int.\ J.\ Mod.\ Phys.\ A {\bf 20}, 2859 (2005)
  [arXiv:hep-th/0307155].
\par
  F.~Ardalan, H.~Arfaei and N.~Sadooghi,
  \textit{On the anomalies and Schwinger terms in noncommutative gauge theories}, To appear in Int. J. Mod.
  Phys. A (2006),  [arXiv:hep-th/0507230].
\bibitem{DNS}
M.~R.~Douglas and N.~A.~Nekrasov, \textit{Noncommutative field theory}, Rev.\ Mod.\ Phys.\  {\bf 73}, 977 (2001)
[arXiv:hep-th/0106048].
\par
R.~J.~Szabo, \textit{Quantum field theory on noncommutative spaces}, Phys.\ Rept.\  {\bf 378}, 207 (2003)
[arXiv:hep-th/0109162].
\bibitem{MRS}
S.~Minwalla, M.~Van Raamsdonk and N.~Seiberg, \textit{Noncommutative perturbative dynamics}, JHEP {\bf 0002},
020 (2000) [arXiv:hep-th/9912072].
\bibitem{Bigatti}
  D.~Bigatti and L.~Susskind,
  \textit{Magnetic fields, branes and noncommutative geometry,}
  Phys.\ Rev.\ D {\bf 62}, 066004 (2000)
  [arXiv:hep-th/9908056].
\bibitem{EV}
E.~V.~Gorbar and V.~A.~Miransky, \textit{Relativistic field theories in a magnetic background as noncommutative
field theories}, Phys.\ Rev.\ D {\bf 70}, 105007 (2004) [arXiv:hep-th/0407219].
\bibitem{ESV}
E.~V.~Gorbar, S.~Homayouni and V.~A.~Miransky, \textit{Chiral dynamics in QED and QCD in a magnetic background
and nonlocal noncommutative field theories}, Phys.\ Rev.\ D {\bf 72}, 065014 (2005) [arXiv:hep-th/0503028].
\bibitem{EMV}
E.~V.~Gorbar, M.~Hashimoto and V.~A.~Miransky, \textit{Nondecoupling phenomena in QED in a magnetic field and
noncommutative QED}, Phys.\ Lett.\ B {\bf 611}, 207 (2005) [arXiv:hep-th/0501135].
\bibitem{Hash}
M.~Hashimoto, \textit{Noncommutativity vs. transversality in QED in a strong magnetic field}, Int.\ J.\ Mod.\
Phys.\ A {\bf 20}, 6307 (2005) [arXiv:hep-th/0507083].
 \bibitem{BK}
  Z.~Bern and D.~A.~Kosower,
  \textit{Efficient calculation of one-loop QCD amplitudes},
  Phys.\ Rev.\ Lett.\  {\bf 66}, 1669 (1991).
\par
  Z.~Bern and D.~A.~Kosower,
  \textit{Color decomposition of one-loop amplitudes in gauge theories},
  Nucl.\ Phys.\ B {\bf 362}, 389 (1991).
\par
  Z.~Bern and D.~A.~Kosower,
  \textit{The Computation of loop amplitudes in gauge theories},
  Nucl.\ Phys.\ B {\bf 379}, 451 (1992).
\bibitem{Strassler}
M.~J.~Strassler, \textit{Field theory without Feynman diagrams: One-loop effective actions}, Nucl.\ Phys.\ B
{\bf 385}, 145 (1992) [arXiv:hep-ph/9205205].
\bibitem{focksch}
M.~G.~Schmidt and C.~Schubert, \textit{On the calculation of effective actions by string methods}, Phys.\ Lett.\
B {\bf 318}, 438 (1993) [arXiv:hep-th/9309055].
\par
J.~W.~van Holten, \textit{Relations between some analytic representations of one-loop scalar integrals}, Z.\
Phys.\ C {\bf 66}, 303 (1995) [arXiv:hep-th/9408027].
\par
M.~Reuter, M.~G.~Schmidt and C.~Schubert, \textit{Constant external fields in gauge theory and the spin 0, 1/2,
1 path integrals}, Annals Phys.\  {\bf 259}, 313 (1997) [arXiv:hep-th/9610191].
\par
W.~Dittrich, \textit{Vacuum polarization using quantum mechanical path integrals}, arXiv:hep-th/\-0005231.
\bibitem{worev}
C.~Schubert, \textit{Perturbative quantum field theory in the string-inspired formalism}, Phys.\ Rept.\  {\bf
355}, 73 (2001) [arXiv:hep-th/0101036], and the references therein.
\bibitem{Gusynin}
V.~P.~Gusynin, V.~A.~Miransky and I.~A.~Shovkovy, \textit{Dimensional reduction and dynamical chiral symmetry
breaking by a magnetic field in (3+1)-dimensions}, Phys.\ Lett.\ B {\bf 349}, 477 (1995) [arXiv:hep-ph/9412257].
\bibitem{Schwinger}
J.~S.~Schwinger, \textit{On gauge invariance and vacuum polarization}, Phys.\ Rev.\  {\bf 82}, 664 (1951).
\bibitem{chodos}
A.~Chodos, K.~Everding and D.~A.~Owen, \textit{QED with a chemical potential: 1. The case of a constant magnetic
field}, Phys.\ Rev.\ D {\bf 42}, 2881 (1990).
\bibitem{Handbook}
M. Abramowitz and I. Stegun, \textit{Handbook of Mathematical Functions }, Dover, 1964.
\bibitem{Zuber}
C. Itzykson and J. Zuber, \textit{Quantum Field Theory}, McGraw-Hill, 1985.




\end{thebibliography}
\end{document}